\documentclass[a4paper]{llncs}

\usepackage[utf8]{inputenc}
\usepackage[T1]{fontenc}
\usepackage{pslatex}
\usepackage[english]{babel}
\usepackage{amsfonts}
\usepackage{amsmath, amssymb}
\usepackage{cancel}
\usepackage{graphicx}
\usepackage{tikz}
\usepackage{txfonts}
\usepackage{hyperref}
\usepackage{url}
\usepackage{fancyhdr}
\usetikzlibrary{positioning}
\usetikzlibrary{decorations.pathreplacing}
\usepackage{subcaption}
\newcommand{\N}{\mathbb{N}}

\newcommand{\F}{\mathbb{F}}

\newtheorem{openproblem}{Open Problem}

\title{Constructing Orthogonal Latin Squares from Linear Cellular Automata}

\author{Luca Mariot\inst{1,2}, Enrico Formenti\inst{2} \and Alberto Leporati\inst{1}}

\institute{Dipartimento di Informatica, Sistemistica e Comunicazione,
  Universit\`{a} degli Studi di Milano-Bicocca, Viale Sarca 336, 20126 Milano,
  Italy \\ \email{\{luca.mariot,alberto.leporati\}@unimib.it} \and
  Laboratoire I3S, Universit\'{e} Nice Sophia Antipolis, 2000 Route des Lucioles,
  06903~Sophia~Antipolis, France \\ \email{\{mariot,enrico.formenti\}@i3s.unice.fr}}

\begin{document}

\maketitle

\begin{abstract}
We undertake an investigation of combinatorial designs engendered by cellular
automata (CA), focusing in particular on orthogonal Latin squares and orthogonal
arrays. The motivation is of cryptographic nature. Indeed, we consider the
problem of employing CA to define threshold secret sharing schemes via
orthogonal Latin squares. We first show how to generate Latin squares through
bipermutive CA. Then, using a characterization based on Sylvester matrices, we
prove that two linear CA induce a pair of orthogonal Latin squares if and only
if the polynomials associated to their local rules are relatively prime.
\end{abstract}


\thispagestyle{fancy}

\section{Introduction}
\emph{Secret sharing schemes} (SSS) are a cryptographic primitive underlying
several protocols such as \emph{secure multiparty computation}~\cite{chaum} and
\emph{generalized oblivious transfer}~\cite{tassa}. The basic scenario addressed
by SSS considers a dealer who wants to share a secret $S$ among a set of $n$
players, so that only certain authorized subsets of players specified in an
access structure may reconstruct $S$. In a $(t,n)$--\emph{threshold scheme}, at
least $t$ out of $n$ players must combine their \emph{shares} in order to
recover $S$, while coalitions with less than $t$ participants learn nothing
about the secret (in an information-theoretic sense).

Recently, a SSS based on cellular automata (CA) has been described
in~\cite{mariot}, where the shares are represented by blocks of a CA
configuration. The main drawback of such proposal is that the access structure
has a \emph{sequential threshold}: in addition to having at least $t$ players,
the shares of an authorized subset must also be adjacent blocks, since they are
used to build a spatially periodic preimage of a CA.

In order to design a CA-based SSS with an unrestricted threshold access
structure, in this paper we take a different perspective that focuses on
\emph{combinatorial designs}. Indeed, it is known that threshold schemes are
equivalent to \emph{orthogonal arrays} (OA), and for $t=2$ the latter correspond
to \emph{mutually orthogonal Latin squares} (MOLS).

The aim of this work is to begin tackling the design of a CA-based threshold
scheme by investigating which CA are able to generate orthogonal Latin
squares. To this end, we first show that every bipermutive cellular automaton of
radius $r$ and length $2m$ induces a Latin square of order $q^m$, where $q$ is
the cardinality of the CA state alphabet and $m$ is any multiple of $2r$. We
then investigate which pairs of bipermutive CA induce orthogonal Latin squares,
by first observing through some experiments that only some pairs of
\emph{linear} CA seem to remain orthogonal upon iteration. We thus prove that
the orthogonality condition holds if and only if the Sylvester matrix built by
juxtaposing the transition matrices of two linear CA is invertible, i.e. if and
only if the polynomials associated to their local rules are relatively
prime. Finally, we show what are the consequences of this result for the design
of CA-based threshold schemes, remarking that the dealer can perform the sharing
phase by evolving a set of linear CA.

The remainder of this paper is organized as follows. Section 2 covers the
preliminary definitions and facts about cellular automata, Latin squares,
orthogonal arrays and secret sharing schemes necessary to describe our
results. Section 3 presents the main contributions of the paper, namely the
proof that a pair of linear CA induce orthogonal Latin squares if and only if
the associated polynomials are coprime. Finally, Section~4 puts the results in
perspective, and discusses some open problems for further research on the topic.

\pagestyle{plain}

\section{Basic Definitions}

\subsection{Cellular Automata}
In this work, we consider one-dimensional CA as \emph{finite compositions of
  functions}, as the next definition summarizes:

\begin{definition}
\label{def:ca}
Let $n$, $r$, $t$ be positive integers such that $t < \left \lfloor \frac{n}{2r}
\right \rfloor$, and let $f: A^{2r+1} \rightarrow A$ be a function of $2r+1$
variables over a finite set $A$ of $q \in \N$ elements. The \emph{cellular
  automaton} (CA) $\langle n, r, t, f\rangle$ is a map $\mathcal{F}: A^n
\rightarrow A^{n-2rt}$ defined by the following composition of functions:
\begin{equation}
\label{eq:comp-ca}
\mathcal{F} = F_{t-1} \circ F_{t-2} \circ \cdots \circ F_1 \circ F_0 \enspace ,
\end{equation}
where for $i \in \{0,\cdots,t-1\}$ function $F_i:A^{n-2ri} \rightarrow
A^{n-2r(i+1)}$ is defined as:
\begin{equation}
F_i(x) =
(f(x_0,\cdots,x_{2r}),f(x_1,\cdots,x_{2r+1}),\cdots,f(x_{n-2r(i+1)-1},\cdots,x_{n-2ri-1})) \enspace
,
\end{equation}
for all $x = (x_0,\cdots,x_{n-2ri-1}) \in A^{n-2ri}$. In particular $n$, $r$ and
$f$ are respectively called the \emph{length}, the \emph{radius} and the
\emph{local rule} of the CA, while for all $i \in \{0,\cdots,t-1\}$ function
$F_i$ is called the \emph{global rule} of the CA at step $i$.
\end{definition}
In some of the results proved in this paper we assume that the state alphabet
$A$ is a \emph{finite field}, i.e. $A = \F_q$ for $q = p^{\alpha}$ where $p$ is
a prime number and $\alpha \in \N$.

A local rule $f: A^{2r+1} \rightarrow A$ is \emph{rightmost permutive}
(respectively, \emph{leftmost permutive}) if, by fixing the value of the first
(respectively, last) $2r$ variables the resulting restriction on the rightmost
(respectively, leftmost) variable is a permutation over $A$. A local rule which
is both leftmost and rightmost permutive is \emph{bipermutive}, and a CA
$\mathcal{F}$ whose local rule is bipermutive is a \emph{bipermutive CA}.

Denoting by $+$ and $\cdot$ respectively sum and multiplication over the finite
field $\F_q$, a local rule $f: \F_q^{2r+1} \rightarrow \F_q$ is \emph{linear} if
there exist $a_0,a_1,\cdots,a_{2r} \in \F_q$ such that
\begin{equation}
\label{eq:lin-rule}
f(x_0,x_1,\cdots,x_{2r}) = a_0x_0 + a_1x_1 + \cdots + a_{2r}x_{2r} \enspace .
\end{equation}
Analogously, a CA $\mathcal{F}$ whose local rule is linear is called a
\emph{linear} (or \emph{additive}) CA. Notice that a linear rule is bipermutive
if and only if both $a_0$ and $a_{2r}$ are not null. The polynomial associated
to a linear rule $f:\F_q^{2r+1}\rightarrow \F_q$ with coefficients
$a_0,\cdots,a_{2r}$ is defined as
\begin{equation}
\label{eq:pol-lin-rule}
p_f(X) = a_0 + a_1X + \cdots + a_{2r}X^{2r} \in \F_q[X] \enspace .
\end{equation}
In a linear CA $\langle n,r,t,f\rangle$ with local rule $f$ defined by the
coefficients $a_0,\cdots,a_{2r} \in \F_q$, the global rule $F_i:\F_q^{n-2ri}
\rightarrow \F_q^{n-2r(i+1)}$ at step $i \in \{0,\cdots,t-1\}$ is a linear
application defined by the following matrix of $n-2r(i+1)$ rows and $n-2ri$
columns:
\begin{equation}
\label{ca-matr}
M_{F_i} = 
\begin{pmatrix}
  a_0    & \cdots & a_{2r} & 0 & \cdots & \cdots & \cdots & \cdots & 0 \\
  0      & a_0    & \cdots  & a_{2r} & 0 & \cdots & \cdots & \cdots & 0 \\
  \vdots & \vdots & \vdots & \ddots  & \vdots & \vdots & \vdots & \ddots & \vdots \\
  0 & \cdots & \cdots & \cdots & \cdots & 0 & a_0 & \cdots & a_{2r} \\
\end{pmatrix} \enspace .
\end{equation}
Thus, the global rule $F_i$ is defined as $F_i(x) = M_{F_i}x^\top$ for all $x
\in \F_q^{n-2r(i+1)}$, and the composition $\mathcal{F}$ corresponds to the
multiplication of the matrices $M_{F_{t-1}} \cdots M_{F_0}$.

Consider now the case where $n = 2rt+1$. The CA $\mathcal{F}$ maps vectors of
$2rt+1$ components to a single element of $A$. We call this particular function
the $t$\emph{--th iterate} of rule $f$, and we denote it by $f^t$. This leads to
the following equivalence:
\begin{lemma}
\label{lm:eq-ca-iter}
Let $\mathcal{F}:A^{n}\rightarrow A^m$ be a $\langle n,r,t,f \rangle$ CA with
local rule $f:A^{2r+1}\rightarrow A$ such that $n = mk$ and $m=2rs$ for $k,s \in
\N_+$, and $t = m(k-1)/2r$. Then, $\mathcal{F}$ is equivalent to the iterated CA
$\langle n,rt, 1, f^t\rangle$ $\mathcal{F}^{(t)}:A^n\rightarrow A^m$, i.e. for
all $x = (x_0,\cdots,x_{n-1}) \in A^n$ it holds that
\begin{equation}
\label{eq:equiv-ca-iter}
\mathcal{F}(x) = \mathcal{F}_t(x) =
(f^t(x_0,\cdots,x_{2rt}),f^t(x_1,\cdots,x_{2rt+1}),\cdots,f^t(x_{n-2rt-1},\cdots,x_{n-1})) \enspace
.
\end{equation}
\end{lemma}
In particular, if $f:\F_q^{2r+1}\rightarrow \F_q$ is linear with associated
polynomial $p_f(X)$, one can show (see e.g.~\cite{ito}) that
$f^t:\F_q^{2rt+1}\rightarrow \F_q$ is linear for all $t \in \N$, and its
polynomial equals
\begin{equation}
\label{eq:pol-power}
p_{f^t}(X) = [p_f(X)]^t \enspace .
\end{equation}
Thus, the coefficients of the iterated linear rule $f^t$ are simply the
coefficients of the polynomial $p_f(X)^t$.

\subsection{Latin Squares, Orthogonal Arrays and Secret Sharing}
We recall only some facts about Latin squares and orthogonal arrays which are
relevant for threshold schemes, following the notation of
Stinson~\cite{stinson}.
\begin{definition}[Latin square]
\label{def:lat-sq}
Let $X$ be a finite set of $v \in \N$ elements. A \emph{Latin square} of order
$v$ over $X$ is a $v \times v$ matrix $L$ with entries from $X$ such that every
row and every column are permutations of $X$. Two Latin squares $L_1$ and $L_2$
of order $v$ defined over $X$ are \emph{orthogonal} if
$(L_1(i_1,j_1),L_2(i_1,j_1)) \neq (L_1(i_2,j_2),L_2(i_2,j_2))$ for all
$(i_1,j_1) \neq (i_2,j_2)$. In other words, $L_1$ and $L_2$ are orthogonal if by
superposing them one obtains all pairs of the Cartesian product $X \times X$. A
collection of $k$ Latin squares $L_1,\cdots,L_k$ of order $v$ which are pairwise
orthogonal is called a set of $k$ \emph{mutually orthogonal Latin squares}
(MOLS).
\end{definition}
\begin{definition}[Orthogonal array]
\label{def:oa}
Let $X$ be a finite set of $v$ elements, and let $t$, $k$ and $\lambda$ be
positive integers such that $2 \le t \le k$. A $t$--$(v,k,\lambda)$
\emph{orthogonal array} ($t$--$(v,k,\lambda)$--$OA$, for short) is a $\lambda
v^t\times k$ rectangular matrix with entries from $X$ such that, for any subset
of $t$ columns, every $t$--uple $(x_1,\cdots,x_t) \in X^t$ occurs in exactly
$\lambda$ rows.
\end{definition}
A $t$--$(v,k,1)$--$OA$ can be used to implement a $(t,n)$--\emph{threshold
  scheme} with $n = k-1$ players $P_1,\cdots,P_{k-1}$ as follows. The dealer
randomly chooses with uniform probability the secret $S$ from the support set
$X$ and a row $A(i,\cdot)$ in the OA such that the last component equals
$S$. Next, for all $j \in \{1,\cdots,k-1\}$ the dealer distributes to player
$P_j$ the share $s_j=A(i,j)$. Since the array is orthogonal with $\lambda=1$,
any subset of $t$ players $P_{j_1},\cdots,P_{j_t}$ can recover the secret, the
reason being that the shares $(s_{j_1},\cdots,s_{j_t})$ form a $t$--uple which
uniquely identifies row $A(i,\cdot)$. Conversely, suppose that $t-1$ players
$P_{i_1},\cdots,P_{i_{t-1}}$ try to determine the secret. Then, the
$(t-1)$--uple $s = (s_{j_1},\cdots,s_{j_{t-1}})$ occurs in the columns
$j_1,\cdots,j_{t-1}$ in $v$ rows of the array. By considering also the last
column, one obtains a $t$--uple $(s_{j_1},\cdots,s_{j_{t-1}},A(i_h,k))$ for all
$1 \le h \le v$. Since $\lambda=1$, it must be the case that all these
$t$--uples are distinct, and thus they must differ in the last component. Hence,
the $v$ rows where the $(t-1)$--uple $(s_{j_1},\cdots,s_{j_{t-1}})$ appears
determine a permutation on the last column, and thus all the values for the
secret are equally likely.

When $t=2$ and $\lambda=1$, the resulting orthogonal array is a $v^2\times k$
matrix in which every pair of columns contains all ordered pairs of symbols from
$X$. In this case, the orthogonal array is simply denoted as $OA(k,v)$, and it
is equivalent to a set of $k-2$ MOLS. As a matter of fact, suppose that
$L_1,\cdots,L_{k-2}$ are $k-2$ MOLS of order $v$. Without loss of generality, we
can assume that $X = \{1,\cdots,v\}$. Then, consider a matrix $A$ of size
$v^2\times k$ defined as follows:
\begin{itemize}
\item The first two columns are filled with all ordered pairs $(i,j) \in X
  \times X$ arranged in lexicographic order.
\item For all $1\le i \le v^2$ and $3\le h \le k$, the entry $(i,h)$ of $A$ is
  defined as 
\begin{equation}
\label{eq:mols-oa-constr}
A(i,h) = L_{h-2}(A(i,1),A(i,2)) \enspace .
\end{equation}
In other words, column $h$ is filled by reading the elements of the Latin square
$L_{h-2}$ from the top left down to the bottom right.
\end{itemize}
The resulting array is a $OA(k,v)$: indeed, let $h_1,h_2$ be two of its columns.
If $h_1=1$ and $h_2=2$ one gets all the ordered pairs of symbols from $X$ in
lexicographic order. If $h_1=1$ (respectively, $h_1=2$) and $h_2\ge 3$, one
obtains all pairs because the $h_1$-th row (respectively, column) of $L_{h_2-2}$
is a permutation over $X$. Finally, for $h_1\ge 3$ and $h_2\ge 3$ one still gets
all ordered pairs since the Latin squares $L_{h_1-2}$ and $L_{h_2-2}$ are
orthogonal. Due to lack of space, we omit the inverse direction from $OA(k,v)$
to $k-2$ MOLS. The reader can find further details about the construction
in~\cite{stinson}.

\section{Main Results}
We begin by showing that any bipermutive cellular automaton of radius $r$ and
length $2m$ induces a Latin square of order $N = q^m$, under the condition that
$m$ is a multiple of $2r$. To this end, we first need some additional notation
and definitions.

Given an alphabet $A$ of $q$ symbols, in what follows we assume that a total
order $\le$ is defined over the set of $m$--uples $A^m$, and that $\phi: A^m
\rightarrow [N]$ is a monotone one-to-one mapping between $A^m$ and $[N] =
\{1,\cdots,q^m\}$, where the order relation on $[N]$ is the usual order on
natural numbers. We denote by $\psi = \phi^{-1}$ the inverse mapping of $\phi$.

The following definition introduces the notion of square associated to a CA:
\begin{definition}
\label{def:square-ca}
Let $m$, $r$ and $t$ be positive integers such that $m = 2rt$, and let $f:
A^{2r+1}\rightarrow A$ be a local rule of radius $r$ over alphabet $A$ with
$|A|=q$. The \emph{square} associated to the CA $\langle 2m, r, t, f \rangle$
with map $\mathcal{F}: A^{2m} \rightarrow A^m$ is the square matrix
$\mathcal{S}_{\mathcal{F}}$ of size $q^m \times q^m$ with entries from $A^m$
defined for all $1 \le i,j \le q^m$ as
\begin{equation}
\label{eq:sq-ca}
\mathcal{S}_{\mathcal{F}}(i,j) = \phi(\mathcal{F}(\psi(i)||\psi(j))) \enspace ,
\end{equation}
where $\psi(i)||\psi(j) \in A^{2m}$ denotes the concatenation of vectors
$\psi(i),\psi(j) \in A^m$.
\end{definition}
Hence, the square $\mathcal{S}_{\mathcal{F}}$ is defined by encoding the first
half of the CA configuration as the row coordinate $i$, the second half as the
column coordinate $j$ and the output of the CA $\mathcal{F}(\psi(i)||\psi(j))$
as the entry in cell $(i,j)$.

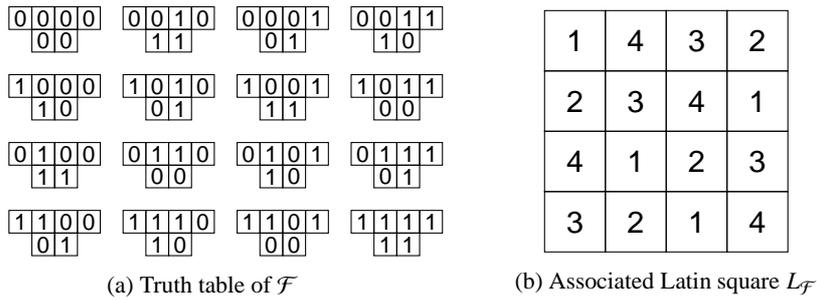
\begin{figure}[b]
\centering
\begin{subfigure}{.5\textwidth}
\centering
\begin{tikzpicture}
[->,auto,node distance=1.5cm,
       empt node/.style={font=\sffamily,inner sep=0pt,minimum size=0pt},
       rect node/.style={rectangle,draw,font=\sffamily,minimum size=0.3cm, inner sep=0pt, outer sep=0pt}]
	
        \node [empt node] (e1) {};
	\node [rect node] (i111) [right=0.5cm of e1] {0};
        \node [rect node] (i112) [right=0cm of i111] {0};
        \node [rect node] (i113) [right=0cm of i112] {0};
        \node [rect node] (i114) [right=0cm of i113] {0};
        \node [rect node] (i115) [below=0cm of i112] {0};
        \node [rect node] (i116) [right=0cm of i115] {0};

        \node [rect node] (i121) [right=0.3cm of i114] {0};
        \node [rect node] (i122) [right=0cm of i121] {0};
        \node [rect node] (i123) [right=0cm of i122] {1};
        \node [rect node] (i124) [right=0cm of i123] {0};
        \node [rect node] (i125) [below=0cm of i122] {1};
        \node [rect node] (i126) [right=0cm of i125] {1};

        \node [rect node] (i131) [right=0.3cm of i124] {0};
        \node [rect node] (i132) [right=0cm of i131] {0};
        \node [rect node] (i133) [right=0cm of i132] {0};
        \node [rect node] (i134) [right=0cm of i133] {1};
        \node [rect node] (i135) [below=0cm of i132] {0};
        \node [rect node] (i136) [right=0cm of i135] {1};

        \node [rect node] (i141) [right=0.3cm of i134] {0};
        \node [rect node] (i142) [right=0cm of i141] {0};
        \node [rect node] (i143) [right=0cm of i142] {1};
        \node [rect node] (i144) [right=0cm of i143] {1};
        \node [rect node] (i145) [below=0cm of i142] {1};
        \node [rect node] (i146) [right=0cm of i145] {0};

	\node [rect node] (i211) [below=0.6cm of i111] {1};
        \node [rect node] (i212) [right=0cm of i211] {0};
        \node [rect node] (i213) [right=0cm of i212] {0};
        \node [rect node] (i214) [right=0cm of i213] {0};
        \node [rect node] (i215) [below=0cm of i212] {1};
        \node [rect node] (i216) [right=0cm of i215] {0};

        \node [rect node] (i221) [right=0.3cm of i214] {1};
        \node [rect node] (i222) [right=0cm of i221] {0};
        \node [rect node] (i223) [right=0cm of i222] {1};
        \node [rect node] (i224) [right=0cm of i223] {0};
        \node [rect node] (i225) [below=0cm of i222] {0};
        \node [rect node] (i226) [right=0cm of i225] {1};

        \node [rect node] (i231) [right=0.3cm of i224] {1};
        \node [rect node] (i232) [right=0cm of i231] {0};
        \node [rect node] (i233) [right=0cm of i232] {0};
        \node [rect node] (i234) [right=0cm of i233] {1};
        \node [rect node] (i235) [below=0cm of i232] {1};
        \node [rect node] (i236) [right=0cm of i235] {1};

        \node [rect node] (i241) [right=0.3cm of i234] {1};
        \node [rect node] (i242) [right=0cm of i241] {0};
        \node [rect node] (i243) [right=0cm of i242] {1};
        \node [rect node] (i244) [right=0cm of i243] {1};
        \node [rect node] (i245) [below=0cm of i242] {0};
        \node [rect node] (i246) [right=0cm of i245] {0};

	\node [rect node] (i311) [below=0.6cm of i211] {0};
        \node [rect node] (i312) [right=0cm of i311] {1};
        \node [rect node] (i313) [right=0cm of i312] {0};
        \node [rect node] (i314) [right=0cm of i313] {0};
        \node [rect node] (i315) [below=0cm of i312] {1};
        \node [rect node] (i316) [right=0cm of i315] {1};

        \node [rect node] (i321) [right=0.3cm of i314] {0};
        \node [rect node] (i322) [right=0cm of i321] {1};
        \node [rect node] (i323) [right=0cm of i322] {1};
        \node [rect node] (i324) [right=0cm of i323] {0};
        \node [rect node] (i325) [below=0cm of i322] {0};
        \node [rect node] (i326) [right=0cm of i325] {0};

        \node [rect node] (i331) [right=0.3cm of i324] {0};
        \node [rect node] (i332) [right=0cm of i331] {1};
        \node [rect node] (i333) [right=0cm of i332] {0};
        \node [rect node] (i334) [right=0cm of i333] {1};
        \node [rect node] (i335) [below=0cm of i332] {1};
        \node [rect node] (i336) [right=0cm of i335] {0};

        \node [rect node] (i341) [right=0.3cm of i334] {0};
        \node [rect node] (i342) [right=0cm of i341] {1};
        \node [rect node] (i343) [right=0cm of i342] {1};
        \node [rect node] (i344) [right=0cm of i343] {1};
        \node [rect node] (i345) [below=0cm of i342] {0};
        \node [rect node] (i346) [right=0cm of i345] {1};

	\node [rect node] (i411) [below=0.6cm of i311] {1};
        \node [rect node] (i412) [right=0cm of i411] {1};
        \node [rect node] (i413) [right=0cm of i412] {0};
        \node [rect node] (i414) [right=0cm of i413] {0};
        \node [rect node] (i415) [below=0cm of i412] {0};
        \node [rect node] (i416) [right=0cm of i415] {1};

        \node [rect node] (i421) [right=0.3cm of i414] {1};
        \node [rect node] (i422) [right=0cm of i421] {1};
        \node [rect node] (i423) [right=0cm of i422] {1};
        \node [rect node] (i424) [right=0cm of i423] {0};
        \node [rect node] (i425) [below=0cm of i422] {1};
        \node [rect node] (i426) [right=0cm of i425] {0};

        \node [rect node] (i431) [right=0.3cm of i424] {1};
        \node [rect node] (i432) [right=0cm of i431] {1};
        \node [rect node] (i433) [right=0cm of i432] {0};
        \node [rect node] (i434) [right=0cm of i433] {1};
        \node [rect node] (i435) [below=0cm of i432] {0};
        \node [rect node] (i436) [right=0cm of i435] {0};

        \node [rect node] (i441) [right=0.3cm of i434] {1};
        \node [rect node] (i442) [right=0cm of i441] {1};
        \node [rect node] (i443) [right=0cm of i442] {1};
        \node [rect node] (i444) [right=0cm of i443] {1};
        \node [rect node] (i445) [below=0cm of i442] {1};
        \node [rect node] (i446) [right=0cm of i445] {1};
	
\end{tikzpicture}
\caption{Truth table of $\mathcal{F}$}
\end{subfigure}%
\begin{subfigure}{.5\textwidth}
\centering
\begin{tikzpicture}
[->,auto,node distance=1.5cm,
       empt node/.style={font=\sffamily,inner sep=0pt,minimum size=0pt},
       rect node/.style={rectangle,draw,font=\sffamily,minimum size=0.8cm, inner sep=0pt, outer sep=0pt}]
	\large
        
	\node [rect node] (s11) {1};
        \node [rect node] (s12) [right=0cm of s11] {4};
        \node [rect node] (s13) [right=0cm of s12] {3};
        \node [rect node] (s14) [right=0cm of s13] {2};

	\node [rect node] (s21) [below=0cm of s11] {2};
        \node [rect node] (s22) [right=0cm of s21] {3};
        \node [rect node] (s23) [right=0cm of s22] {4};
        \node [rect node] (s24) [right=0cm of s23] {1};

	\node [rect node] (s31) [below=0cm of s21] {4};
        \node [rect node] (s32) [right=0cm of s31] {1};
        \node [rect node] (s33) [right=0cm of s32] {2};
        \node [rect node] (s34) [right=0cm of s33] {3};

	\node [rect node] (s41) [below=0cm of s31] {3};
        \node [rect node] (s42) [right=0cm of s41] {2};
        \node [rect node] (s43) [right=0cm of s42] {1};
        \node [rect node] (s44) [right=0cm of s43] {4};
	
\end{tikzpicture}
\caption{Associated Latin square $L_{\mathcal{F}}$}
\end{subfigure}%
\caption{Example of Latin square of order $4$ induced by rule 150. Mapping
  $\phi$ is defined as $\phi(00) \mapsto 1$, $\phi(10) \mapsto 2$, $\phi(01)
  \mapsto 3$, $\phi(11) \mapsto 4$.}
\label{fig:r150-sq}
\end{figure}

The next lemma shows that fixing the leftmost or rightmost $2r$ input variables
in the global rules of a bipermutive CA yields a permutation between the
remaining variables and the output:
\begin{lemma}[\cite{mariot}]
\label{lm:restr-perm}
Let $\mathcal{F}: A^n \rightarrow A^{n-2rt}$ be a bipermutive CA $\langle n, r,
t, f \rangle$ defined by local rule $f: A^{2r+1} \rightarrow A$, and let
$F_i:A^{n-2ri} \rightarrow A^{n-2r(i+1)}$ be its global rule at step $i \in
\{0,\cdots,t-1\}$. Then, by fixing at least $d\ge2r$ leftmost or rightmost
variables in $x \in A^{n-2ri}$ to the values
$\tilde{x}=(\tilde{x}_0,\cdots,\tilde{x}_{d-1})$, the resulting restriction
$F_i|_{\tilde{x}}: A^{n-2r(i+1)} \rightarrow A^{n-2r(i+1)}$ is a permutation.
\end{lemma}
On account of Lemma~\ref{lm:restr-perm}, we now prove that the squares
associated to bipermutive CA are in fact Latin squares. The proof follows the
argument laid out in Lemma 2 of~\cite{mariot}.
\begin{lemma}
\label{lm:lat-sq-bip-ca}
Let $f:A^{2r+1} \rightarrow A$ be a bipermutive local rule defined over $A$ with
$|A|=q$, and let $m = 2rt$ where $t \in \N$. Then, the square $L_{\mathcal{F}}$
associated to the bipermutive CA $\langle 2m, r, t, f \rangle$ $\mathcal{F}:
A^{2m} \rightarrow A^m$ is a Latin square of order $q^m$ over $X =
\{1,\cdots,q^m\}$.
\begin{proof}
Let $i \in \{1,\cdots,q^m\}$ be a row of $L_{\mathcal{F}}$, and let $\psi(i) = x
= (x_0,\cdots,x_{m-1}) \in A^m$ be the vector associated to $i$ with respect to
the total order $\le$ on $A^m$. Consider now a vector $c_0 \in A^{2m}$ whose
first $m$ coordinates coincide with $x$, and let $c_1 = F_0(c_0)$ be the image
of $c$ under the global rule $F_0$. Then, by Lemma~\ref{lm:restr-perm} there is
a permutation $\pi_0:A^m \rightarrow A^m$ between the rightmost $m$ variables of
$c_0$ and the rightmost $m$ ones of $c_1$. Likewise, since the leftmost $m-2r$
coordinates of $c_1$ are determined by applying the restriction of $F_0$ to $x$,
it follows that there exists a permutation $\pi_1:A^m \rightarrow A^m$ between
the rightmost $m$ variables of $c_1$ and the rightmost $m$ ones of $c_2 =
F_1(c_1)$. More in general, since $m$ is a multiple of $2r$, for all steps $i
\in \{2,\cdots, t-1\}$ there are always at least $2r$ leftmost variables of
$c_{i-1}$ determined, and thus by Lemma~\ref{lm:restr-perm} there is a
permutation $\pi_i: A^m \rightarrow A^m$ between the rightmost $m$ variables of
$c_{i-1} = F_{i-1}(c_{i-2})$ and the rightmost $m$ variables of $c_i =
F_i(c_{i-1})$. Consequently, there exists a permutation $\pi: A^m \rightarrow
A^m$ between the rightmost $m$ variables of $c_0$ and the output value of
$\mathcal{F}(c_0)$, defined as:
\begin{equation}
\label{eq:comp-perm}
\pi = \pi_{t-1} \circ \pi_{t-2} \circ \cdots \circ \pi_1 \circ \pi_0 \enspace .
\end{equation}
For all $q^m$ choices of the rightmost $m$ variables of $c_0$, the values at
$L_{\mathcal{F}}(i,\cdot)$ are determined by computing
$\phi(\mathcal{F}(c_0))$. As a consequence, the $i$-th row of $L_{\mathcal{F}}$
is a permutation of $X = \{1,\cdots,q^m\}$. A symmetric argument holds when
considering a column $j$ of $L_{\mathcal{F}}$ with $1\le j \le q^m$, which fixes
the rightmost $m$ variables of $\mathcal{F}$ to the value $\psi(j)$. Hence,
every column of $L_{\mathcal{F}}$ is also a permutation of $X$, and thus
$L_{\mathcal{F}}$ is a Latin square of order $q^m$.\qed
\end{proof}
\end{lemma}
As an example, for $A = \F_2$ and radius $r=1$, Figure~\ref{fig:r150-sq} reports
the Latin square $L_{\mathcal{F}}$ the bipermutive CA $\mathcal{F}: \F_2^4
\rightarrow \F_2^2$ with rule $150$, defined as $f_{150}(x_0,x_1,x_2) = x_0
\oplus x_1 \oplus x_2$.

We now aim at characterizing pairs of CA which generate orthogonal Latin
squares. For alphabet $A=\F_2$ and radius $r=1$ there exist only two bipermutive
rules up to complementation and reflection, which are rule 150 and rule 90, the
latter defined as $f_{90}(x_0,x_1,x_2) = x_0 \oplus x_2$. Both rules are linear,
and for length $n = 4$ their associated Latin squares are orthogonal, as shown
in Figure~\ref{fig:r150-90}.
\begin{figure}[t]
\centering
\begin{subfigure}{.3\textwidth}
\centering
\begin{tikzpicture}
[->,auto,node distance=1.5cm,
       empt node/.style={font=\sffamily,inner sep=0pt,minimum size=0pt},
       rect node/.style={rectangle,draw,font=\sffamily,minimum size=0.7cm, inner sep=0pt, outer sep=0pt}]

	\node [rect node] (s11) {1};
        \node [rect node] (s12) [right=0cm of s11] {4};
        \node [rect node] (s13) [right=0cm of s12] {3};
        \node [rect node] (s14) [right=0cm of s13] {2};

	\node [rect node] (s21) [below=0cm of s11] {2};
        \node [rect node] (s22) [right=0cm of s21] {3};
        \node [rect node] (s23) [right=0cm of s22] {4};
        \node [rect node] (s24) [right=0cm of s23] {1};

	\node [rect node] (s31) [below=0cm of s21] {4};
        \node [rect node] (s32) [right=0cm of s31] {1};
        \node [rect node] (s33) [right=0cm of s32] {2};
        \node [rect node] (s34) [right=0cm of s33] {3};

	\node [rect node] (s41) [below=0cm of s31] {3};
        \node [rect node] (s42) [right=0cm of s41] {2};
        \node [rect node] (s43) [right=0cm of s42] {1};
        \node [rect node] (s44) [right=0cm of s43] {4};
	
\end{tikzpicture}
\caption{Latin square of rule 150}
\end{subfigure}%
\begin{subfigure}{.3\textwidth}
\centering
\begin{tikzpicture}
[->,auto,node distance=1.5cm,
       empt node/.style={font=\sffamily,inner sep=0pt,minimum size=0pt},
       rect node/.style={rectangle,draw,font=\sffamily,minimum size=0.7cm, inner sep=0pt, outer sep=0pt}]

	\node [rect node] (s11) {1};
        \node [rect node] (s12) [right=0cm of s11] {2};
        \node [rect node] (s13) [right=0cm of s12] {3};
        \node [rect node] (s14) [right=0cm of s13] {4};

	\node [rect node] (s21) [below=0cm of s11] {2};
        \node [rect node] (s22) [right=0cm of s21] {1};
        \node [rect node] (s23) [right=0cm of s22] {4};
        \node [rect node] (s24) [right=0cm of s23] {3};

	\node [rect node] (s31) [below=0cm of s21] {3};
        \node [rect node] (s32) [right=0cm of s31] {4};
        \node [rect node] (s33) [right=0cm of s32] {1};
        \node [rect node] (s34) [right=0cm of s33] {2};

	\node [rect node] (s41) [below=0cm of s31] {4};
        \node [rect node] (s42) [right=0cm of s41] {3};
        \node [rect node] (s43) [right=0cm of s42] {2};
        \node [rect node] (s44) [right=0cm of s43] {1};
	
\end{tikzpicture}
\caption{Latin square of rule 90}
\end{subfigure}%
\begin{subfigure}{.3\textwidth}
\centering
\begin{tikzpicture}
[->,auto,node distance=1.5cm,
       empt node/.style={font=\sffamily,inner sep=0pt,minimum size=0pt},
       rect node/.style={rectangle,draw,font=\sffamily,minimum size=0.7cm, inner sep=0pt, outer sep=0pt}]

	\node [rect node] (s11) {1,1};
        \node [rect node] (s12) [right=0cm of s11] {4,2};
        \node [rect node] (s13) [right=0cm of s12] {3,3};
        \node [rect node] (s14) [right=0cm of s13] {2,4};

	\node [rect node] (s21) [below=0cm of s11] {2,2};
        \node [rect node] (s22) [right=0cm of s21] {3,1};
        \node [rect node] (s23) [right=0cm of s22] {4,4};
        \node [rect node] (s24) [right=0cm of s23] {1,3};

	\node [rect node] (s31) [below=0cm of s21] {4,3};
        \node [rect node] (s32) [right=0cm of s31] {1,4};
        \node [rect node] (s33) [right=0cm of s32] {2,1};
        \node [rect node] (s34) [right=0cm of s33] {3,2};

	\node [rect node] (s41) [below=0cm of s31] {3,4};
        \node [rect node] (s42) [right=0cm of s41] {2,3};
        \node [rect node] (s43) [right=0cm of s42] {1,2};
        \node [rect node] (s44) [right=0cm of s43] {4,1};
	
\end{tikzpicture}
\caption{Superposed square}
\end{subfigure}%
\caption{Orthogonal Latin squares generated by bipermutive CA with rule 150 and
  90.}
\label{fig:r150-90}
\end{figure}
For $r=2$ and length $2m = 8$, a computer search among all $256$ bipermutive
rules of radius $2$ yields $426$ pairs of CA which generate orthogonal Latin
squares of order $2^4 = 16$, among which are both linear and nonlinear
rules. However, for length $2m = 16$ only $21$ pairs of linear rules still
generate orthogonal Latin squares of order $2^8 = 256$. For this reason, we
narrowed our investigation only to linear rules. When $m = 2r$, the following
result gives a necessary and sufficient condition on the CA matrices:

\begin{lemma}
\label{lm:ort-ls-4r}
Let $\mathcal{F}:\F_q^{4r}\rightarrow \F_q^{2r}$ and $\mathcal{G}:\F_q^{4r}
\rightarrow \F_q^{2r}$ be linear CA of radius $r$, respectively with linear
rules $f(x_0,\cdots,x_{2r}) = a_0x_0 + \cdots a_{2r}x_{2r}$ and
$g(x_0,\cdots,x_{2r}) = b_0x_0 + \cdots b_{2r}x_{2r}$, where
$a_0,b_0,a_{2r},b_{2r}\neq 0$. Additionally, let $M_{\mathcal{F}}$ and
$M_{\mathcal{G}}$ be the $2r\times 4r$ matrices associated to the global rules
$F_0 = \mathcal{F}$ and $G_0 = \mathcal{G}$ respectively, and define the $4r
\times 4r$ matrix $M$ as
\begin{equation}
\label{eq:sylv-matr}
M = \begin{pmatrix} M_{\mathcal{F}} \\ M_{\mathcal{G}} \end{pmatrix} =
\begin{pmatrix}
  a_0    & \cdots & a_{2r} & 0 & \cdots & \cdots & \cdots & \cdots & 0 \\
  0      & a_0    & \cdots  & a_{2r} & 0 & \cdots & \cdots & \cdots & 0 \\
  \vdots & \vdots & \vdots & \ddots  & \vdots & \vdots & \vdots & \ddots & \vdots \\
  0 & \cdots & \cdots & \cdots & \cdots & 0 & a_0 & \cdots & a_{2r} \\
  b_0    & \cdots & b_{2r} & 0 & \cdots & \cdots & \cdots & \cdots & 0 \\
  0      & b_0    & \cdots  & b_{2r} & 0 & \cdots & \cdots & \cdots & 0 \\
  \vdots & \vdots & \vdots & \ddots  & \vdots & \vdots & \vdots & \ddots & \vdots \\
  0 & \cdots & \cdots & \cdots & \cdots & 0 & b_0 & \cdots & b_{2r} \\
\end{pmatrix} \enspace .
\end{equation} 
Then, the Latin squares $L_{\mathcal{F}}$ and $L_{\mathcal{G}}$ generated by
$\mathcal{F}$ and $\mathcal{G}$ are orthogonal if and only if the determinant of
$M$ over $\F_q$ is not null.
\begin{proof}
Denote by $z=x||y$ the concatenation of vectors $x$ and $y$. We have to show
that the function $\mathcal{H}: \F_q^{2r}\times \F_q^{2r} \rightarrow
\F_q^{2r}\times\F_q^{2r}$, defined for all $(x,y) \in \F_q^{2r}\times \F_q^{2r}$
as
\begin{equation}
\label{eq:map-h}
\mathcal{H}(x,y) = (\mathcal{F}(z), \mathcal{G}(z)) = (\tilde{x}, \tilde{y})
\end{equation}
is bijective. Let us rewrite Equation~\eqref{eq:map-h} as a system of two
equations:
\begin{equation}
\label{eq:system-1}
\begin{cases}
\mathcal{F}(z) = M_{\mathcal{F}}z^T = \tilde{x} \\
\mathcal{G}(z) = M_{\mathcal{G}}z^T = \tilde{y}
\end{cases}
\end{equation}
As $M$ consists of the juxtaposition of $M_{\mathcal{F}}$ and $M_{\mathcal{G}}$,
Equation~\eqref{eq:system-1} defines a linear system in $4r$ equations and $4r$
unknowns with associated matrix $M$. Thus, we have that $\mathcal{H}(x,y) =
Mz^T$, and $\mathcal{H}$ is bijective if and only if the determinant of $M$ is
not null. \qed
\end{proof}
\end{lemma}
Remark that matrix $M$ in Equation~\eqref{eq:sylv-matr} is a \emph{Sylvester
  matrix}, and its determinant is the \emph{resultant} of the two polynomials
$p_f(X)$ and $p_g(X)$ associated to $f$ and $g$ respectively. The resultant of
two polynomials is nonzero if and only if they are relatively prime
(see~\cite{lidl}). Clearly, if $p_f(X)$ and $p_g(X)$ are relatively prime, then
for any $t \in \N$ their powers $p_f(X)^t$ and $p_g(X)^t$ will be relatively
prime as well. Additionally, $p_f(X)^t$ and $p_g(X)^t$ are the polynomials of
the $t$-th iterates $f^t$ and $g^t$. By Lemma~\ref{lm:eq-ca-iter}, the linear CA
$\langle 2m, r,t,f\rangle$ and $\langle 2m, r, t, g\rangle$ with maps
$\mathcal{F},\mathcal{G}:A^{2m}\rightarrow A^m$ are equivalent to the linear CA
$\langle 2m, rt, 1, f^t\rangle$ and $\langle 2m, rt, 1, g^t\rangle$ with maps
$\mathcal{F}_t,\mathcal{G}_t:A^{2m}\rightarrow A^m$ for any multiple $m \in \N$
of $2r$. We thus have the following result:
\begin{theorem}
\label{th:lat-sq-ca}
Let $f,g:\F_q^{2r+1}\rightarrow \F_q$ be linear bipermutive rules of radius
$r\in \N$. Then, for any $t \in \N$ and $m = 2rt$, the squares $L_{\mathcal{F}}$
and $L_{\mathcal{G}}$ of order $q^m$ respectively associated to the linear CA
$\langle 2m,r,t,f \rangle$ $\mathcal{F}:\F_q^{2m} \rightarrow \F_q^m$ and the
linear CA $\langle 2m,r,t,g \rangle$ $\mathcal{G}:\F_q^{2m} \rightarrow \F_q^m$
are orthogonal if and only if the polynomials $p_f(X)$ and $p_g(X)$ are
relatively prime.
\end{theorem}

\section{Conclusions and Perspectives}
By Theorem~\ref{th:lat-sq-ca}, one can generate a set of $n$ MOLS of order $q^m$
through linear CA of radius $r$ by finding $n$ pairwise relatively prime
polynomials of degree $2r$, where $2r|m$. The problem of counting the number of
pairs of relatively prime polynomials over finite fields has been considered in
several works (see for example~\cite{reifegerste,benjamin,hou}). However, notice
that determining the number of pairs of linear CA inducing orthogonal Latin
squares entails counting only specific pairs of polynomials, namely those whose
constant term is not null. This is due to the requirement that the CA local
rules must be bipermutive. As far as the authors know, this particular version
of the counting problem for relatively prime polynomials has not been addressed
in the literature, for which reason we formalize it below as an open problem for
future investigation:

\begin{openproblem}
Let $f,g \in \F_q[x]$ be defined as follows:
\begin{align*}
f(x) &= a + a_1x + \cdots + a_{n-1}x^{n-1} + x^n \enspace ,\\
g(x) &= b + b_1x + \cdots + b_{n-1}x^{n-1} + x^n \enspace ,
\end{align*}
where $a\neq 0$ and $b \neq 0$. Let $P_n^{a,b}$ be the set of pairs $(f,g)$ of
all such polynomials, and define $C_n^{a,b}$ as
\begin{displaymath}
C_{n}^{a,b} = \{(f,g) \in P_{n}^{a,b}: \gcd(f,g) = 1\}
\end{displaymath}
Then, what is the cardinality of $C_n^{a,b}$?
\end{openproblem}

Given the equivalence between MOLS and OA, Theorem~\ref{th:lat-sq-ca} also gives
some additional insights on how to design a CA-based secret sharing scheme with
threshold $t=2$. In particular, suppose that the secret $S$ is a vector of
$\F_q^m$, and there are $n$ players $P_1,\cdots,P_n$. The dealer picks $n$
relatively prime polynomials of degree $2r$, and builds the corresponding linear
rules $f_1,\cdots,f_n$ of radius $r$. For practical purposes, the dealer could
settle for $n$ irreducible polynomials, for which there exist several efficient
generation algorithms in the literature (see for
instance~\cite{shoup}). Successively, the dealer concatenates the secret $S$
with a random vector $R \in \F_q^m$, thus obtaining a configuration $C \in
\F_q^{2m}$ of length $2m$. Adopting the point of view of OA, this step
corresponds to the phase where the dealer chooses one of the rows of the array
whose first component is the secret. In order to determine the remaining
components of the row, and thus the shares to distribute to the players, for all
$i \in \{1,\cdots,n\}$ the dealer evolves the CA $\mathcal{F}_i$ with rule $f_i$
starting from configuration $C$. The value $B_i=\mathcal{F}_i(C)$ constitutes
the share of player $P_i$.

For the recovery phase, suppose that two players $P_i$ and $P_j$ want to
determine the secret. Since the orthogonal array is assumed to be public, both
$P_i$ and $P_j$ know the CA linear rules $f_i$ and $f_j$ used by the dealer to
compute their shares. Hence, they invert the corresponding Sylvester matrix, and
multiply it for the concatenated vector $(B_i||B_j)$. By
Lemma~\ref{lm:ort-ls-4r}, the result of this multiplication will be the
concatenation of secret $S$ and random vector $R$.

\end{document}